\documentclass[aps,pre,reprint,showpacs,notitlepage,onecolumn]{revtex4-1}
\usepackage{epstopdf,epsfig,graphics,amssymb,amsmath,subeqnarray,color,bm,mathtools,xcolor,float,graphicx,dcolumn,lipsum}
\usepackage[toc,page]{appendix}

\numberwithin{equation}{section}
\usepackage[colorlinks=true,linkcolor=blue]{hyperref}

\def \be{\mathbf{e}}

\def \bf{\mathbf{f}}

\def \bn{\mathbf{n}}

\def \bt{\mathbf{t}}
\def \bu{\mathbf{u}}

\def \bx{\mathbf{x}}

\def \buh{\hat{\bu}}

\definecolor{dgreen}{rgb}{0.0, 0.4, 0.0}
\def \sgn{\text{sgn}}
\def \etal{\textit{et al.~}}

\DeclareMathOperator{\sech}{sech}

\begin{document}

\title{Propulsion via flexible flapping in granular media}
\author{Zhiwei Peng}
\affiliation{Beijing Computational Science Research Center, Beijing 100193, China}
\author{Yang Ding}
\affiliation{Beijing Computational Science Research Center, Beijing 100193, China}
\author{Kyle Pietrzyk}
\affiliation{Department of Mechanical Engineering,
Santa Clara University, Santa Clara, California 95053, USA}
\author{Gwynn J. Elfring}
\affiliation{Department of Mechanical Engineering, University of British Columbia, Vancouver, British Columbia V6T 1Z4, Canada}
\author{On Shun Pak}
\affiliation{Department of Mechanical Engineering,
Santa Clara University, Santa Clara, California 95053, USA}
\date{\today}


\begin{abstract}
Biological locomotion in nature is often achieved by the interaction between a flexible body and its surrounding medium. The interaction of a flexible body with granular media is less understood compared with viscous fluids partially due to its complex rheological properties. In this work we explore the effect of flexibility on granular propulsion by considering a simple mechanical model where a rigid rod is connected to a torsional spring that is under a displacement actuation using a granular resistive force theory. Through a combined numerical and asymptotic investigation, we characterize the propulsive dynamics of such a flexible flapper in relation to the actuation amplitude and spring stiffness and compare these dynamics with those observed in a viscous fluid. In addition, we demonstrate that the maximum possible propulsive force can be obtained in the steady propulsion limit with a finite spring stiffness and large actuation amplitude. These results may apply to the development of synthetic locomotive systems that exploit flexibility to move through complex terrestrial media.
\end{abstract}

\maketitle
\section{Introduction}
\label{sec:intro}

Biological locomotion in nature, such as the swimming of microorganisms in fluids \cite{childress1981mechanics, Lauga2009rev} or the flying of birds in air \cite{Wu2011}, is often achieved by the interaction between a flexible body and its surrounding environment. Despite the diversity in their material composition or internal anatomy, organisms often exhibit bending deformation during motion in fluids \cite{lucas2014bending}. Due to its fundamental significance in understanding natural locomotion as well as its potential applications in the design of artificial locomotive systems, there has been considerable interests in how to exploit flexibility for propulsion enhancement.

For locomotion at high Reynolds numbers, such as swimming of fish and flying of birds, various studies have shown that flexibility can lead to improvements in propulsive performance \cite{katz1978hydrodynamic,alben2008optimal,Michelin2009,spagnolie2010,Thiria2010,Wu2011,lucas2014bending,moore2015torsional,Tytell2016prf}. In particular, Moore \cite{moore2015torsional} has shown that optimal propulsion can be achieved by having a localized flexibility arrangement at the front of the wing via a torsional spring for small amplitude flapping.

In the low Reynolds number regime where viscous forces dominate and inertial forces are negligible, the role of flexibility in propulsion can be significant. A simple illustration is the case of a filament immersed in a viscous fluid that is driven at one end. No net propulsive thrust can be generated when the filament is rigid as constrained by Purcell's scallop theorem \cite{purcell1977life}, because the motion is reciprocal (exhibiting time-reversal symmetry); however, by introducing flexibility, bending deformations due to elastohydrodynamic interactions break the time-reversal symmetry and lead to propulsion \cite{Wiggins1998,WigginsPRL1998}. As a mechanical model emulating flagellar locomotion, a flexible slender filament that is driven by different boundary actuation mechanisms or allowed to swim freely has been studied both experimentally and theoretically \cite{Yu2006,Lauga2007,Evans2010}. For instance, the optimal uniform stiffness can be determined to maximize the propulsive force or swimming speed for a given driven mechanism. Meanwhile, various flexible artificial propellers have been developed in recent years such as nanowires and DNA linked with magnetic beads to understand swimming behaviors at small scales \cite{dreyfus2005microscopic,pak2011SoftMatter,williams2014Nature,Maier2016NANO}.

Compared to swimming and flying in Newtonian fluids, locomotion in terrestrial substances such as granular media is less understood mainly due to their complex rheological properties \cite{zhang2014effective,Goldman2014}. Nevertheless, there has been a developing interest in modeling biological motion on/within granular media such as the subsurface undulatory swimming of sandfish lizards \cite{maladen2009undulatory,maladen2011undulatory}. Similar to the resistive force theory in viscous fluids, Maladen \etal \cite{maladen2009undulatory} developed an empirical resistive force theory in dry granular media which was shown effective in describing the dynamics of slender body locomotion \cite{zhang2014effective}. Employing this granular resistive force theory,  we recently characterized the complex swimming behavior of finite slender swimmers by prescribing a deformation waveform emulating those of sandfish lizards \cite{peng2016pof}. Despite this progress, the role of flexibility on locomotion in granular media remains largely unexplored. In this paper we consider the propulsion generated by a flexible body (flapper) deforming due to interaction with a surrounding granular medium.
 
As a minimal model to understand the role of flexibility in biological propulsion, the arrangement of a rigid rod (or wing) coupled with a torsional spring, which represents localized flexibility, has been studied in Newtonian fluids across Reynolds numbers \cite{spagnolie2010,moore2014analytical,moore2015torsional,Or14PRE,Or16PRE,Giraldi2016,chen2016jfm}. In this work, we consider a torsional spring flapper as a simple mechanism which employs flexibility to move in complex terrestrial media (Fig.~\ref{fig:schematic}). We characterize its propulsive performance arising from the interplay between elastic and granular forces and investigate strategies to maximize the propulsive force generated.

The paper is organized as follows. We formulate our problem and contrast the viscous and granular resistive force models in Sec.~\ref{sec:setup}. Propulsive characteristics of the torsional spring model in a viscous fluid are first discussed and then compared and contrasted with propulsive forces generated in granular media (Sec.~\ref{sec:character_prop}). In Sec.~\ref{sec:asymp}, we discuss different asymptotic regimes for the granular case. We conclude our work with remarks in Sec.~\ref{sec:conclude}.

\section{Setup}
\label{sec:setup}

\subsection{Flexible flapper}
\label{subsec:torsion}
\begin{figure}[t]
\centering
\includegraphics[scale=1]{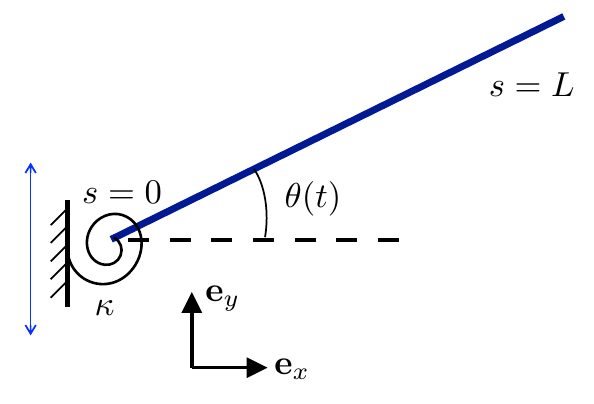}
\caption{\label{fig:schematic}Schematic diagram of a displacement driven rigid filament via a torsional spring connection. }
\end{figure}

We consider a rigid filament of length $L$ and radius $r$ connected to a torsional spring with a spring constant $\kappa$ as illustrated in Fig.~\ref{fig:schematic}. The filament is slender, $r/L\ll 1$, and is harmonically actuated in the horizontal plane (perpendicular to the direction of gravity) at the torsional spring end. Under this actuation, the motion of the filament is confined in the $x$-$y$ plane spanned by the basis vectors $\be_x$ and $\be_y$ while $\be_z = \be_x\times\be_y$. We define the position of a material point on the filament at location $s$ and time $t$ as $\bx(s,t)$, where $s\in[0,L]$ is the arclength along the centerline of the filament. The local unit tangent and normal vectors are given by $\bt = \bx_s = \cos\theta\be_x+ \sin\theta\be_y$ and $\bn = \be_z\times \bt$, where $\theta(t)$ is the angle between $\bt$ and $\be_x$ and the subscript $s$ denotes differentiation with respect to arclength.

At the actuation end, we specify a periodic displacement $\bx_0 = y_0\sin\omega t\be_y$, where $\omega$ is the frequency and $y_0$ the amplitude of actuation. Since the filament is rigid, we may describe the position of points along the centerline of the filament by $\bx = \bx_0 + s \bt$ for $s\in[0,L]$ and then the velocity of those points is $\bu = \dot{\bx}_0 + s \dot{\theta}\bn$, where a dot denotes time derivative.

The governing equation for $\theta(t)$ is then given by a balance of the torque due to resistive forces on the filament and the elastic torque from the torsional spring,
\begin{align}\label{eq:goveq}
\be_z\cdot\int_0^L (\bx-\bx_0)\times\bf~ \text{d} s - \kappa \theta=0,
\end{align}
where $\bf$ is given by the viscous force per unit length ($\bf_{\text{v}}$) when the surrounding environment is a viscous fluid, and by the granular force per unit length ($\bf_{\text{g}}$) if it is a granular medium. Finally, the propulsive force generated by the motion of the filament is given by
\begin{align}
F = -\be_x\cdot\left< \int_0^L\bf \text{d} s \right>,
\end{align}
where the brackets indicate time-averaging over a period of actuation.

\subsection{Resistive force theory}
\label{subsec:dyn}
In a Newtonian fluid, the viscous force experienced by a slender filament at low Reynolds numbers can be approximated by a viscous resistive force theory (RFT) wherein the viscous force per unit length along the body is given by
\begin{align}\label{eq:viscousRFT}
\bf_\text{v}(s,t) = -\left( \xi_\parallel\bt\bt+\xi_\perp \bn\bn   \right)\cdot\bu.
\end{align}
In other words, the viscous force density is linearly related to the local velocity vector of the filament with a tangential ($\xi_\parallel$) and normal ($\xi_\perp$) resistive coefficients. We define the resistance ratio as $\gamma = \xi_\perp /\xi_\parallel$. In the limit of an infinitely slender filament, $L/r \to \infty$, the resistance ratio $\gamma=2$ \cite{GRAY802,lighthillRev}. 

Similar to the overdamped dynamics in fluids at low Reynolds numbers, we consider the case where the granular medium is in a regime where grain-grain and grain-body frictional forces dominate and inertial forces are negligible. In this quasi-static regime, forces exerted by the surrounding granular materials on the filament can be approximated by the so-called granular RFT \cite{maladen2009undulatory}, which has been shown effective in modeling locomotion of slender bodies in granular media \cite{maladen2011undulatory,zhang2014effective,peng2016pof}. In a recent work by Askari and Kamrin \cite{askari2016intrusion}, the effectiveness of RFT in capturing the essential dynamics of a dry granular medium is shown to be a result of the local frictional yielding and no cohesion.

The resistive force per unit length experienced by a slender rod is given by
\begin{align}\label{eq:RFTGM}
\bf_\text{g}(s,t)= -C_\parallel \buh\cdot\bt\bt-C_\perp(\buh-\buh\cdot\bt\bt),
\end{align}
where $\hat{\bu} = \bu/ \lVert\bu \rVert$ is the direction of the velocity and the tangential $C_\parallel$ and normal $C_\perp$ resistive coefficients are related to rheological parameters $C_S, C_F, \gamma_0$ of the granular medium and the radius $r$ of the rod \cite{maladen2009undulatory}:
\begin{align}\label{eq:RFTco}
&C_\parallel = 2rC_F,\\
&C_\perp = C_\parallel \left(1+\frac{C_S}{C_F\sqrt{\tan^2\gamma_0+ (\hat{\bu}\cdot\bn)^2}} \right).
\end{align}
We note that drag anisotropy ($C_\perp > C_\parallel$) is present in granular media and that the normal resistive coefficient depends on the orientation of the rod with respect to the direction of motion while the tangential coefficient is a constant. As opposed to resistive force theory in viscous fluids, the resistive force in granular media does not depend on the magnitude of velocity due to its frictional nature. In this work, numerical values of the aforementioned rheological parameters are adopted from the work by Maladen \etal \cite{maladen2009undulatory} for a loosely packed granular medium. 

\subsection{Dimensionless equations}
\label{subsec:dimless}
We non-dimensionalize the governing equation \eqref{eq:goveq} with respect to the time scale $\omega^{-1}$ and length scale $L$. In granular media, the two resulting dimensionless numbers characterizing the dynamics of the filament under actuation are the dimensionless amplitude $b$ and dimensionless spring constant $K_{\text{g}}$, where
\begin{align}
b = \frac{y_0}{L},\quad K_\text{g}= \frac{\kappa}{C_\parallel L^2}.
\end{align}
The dimensionless spring constant $K_\text{g}$ compares the magnitude of characteristic elastic forces $\kappa/L$ and granular resistive forces $C_\parallel L$. We may write the dimensionless governing equation as
\begin{align}
\int_0^1 s(\hat{\bu}\cdot\bn) \left( 1+\frac{C_S}{C_F\sqrt{\tan^2\gamma_0+(\hat{\bu}\cdot\bn)^2}}  \right)\text{d} s+ K_\text{g} \theta=0. \label{eqn:governing}
\end{align}
The absence of $\omega$ in the dimensionless groups indicates that the dynamics of this system are independent of the actuation frequency. The corresponding dimensionless groups and equations in the case of a viscous fluid are given in Sec.~\ref{subsec:viscous}.

\begin{figure}[!tb]
\centering
\includegraphics[width = 0.45\textwidth]{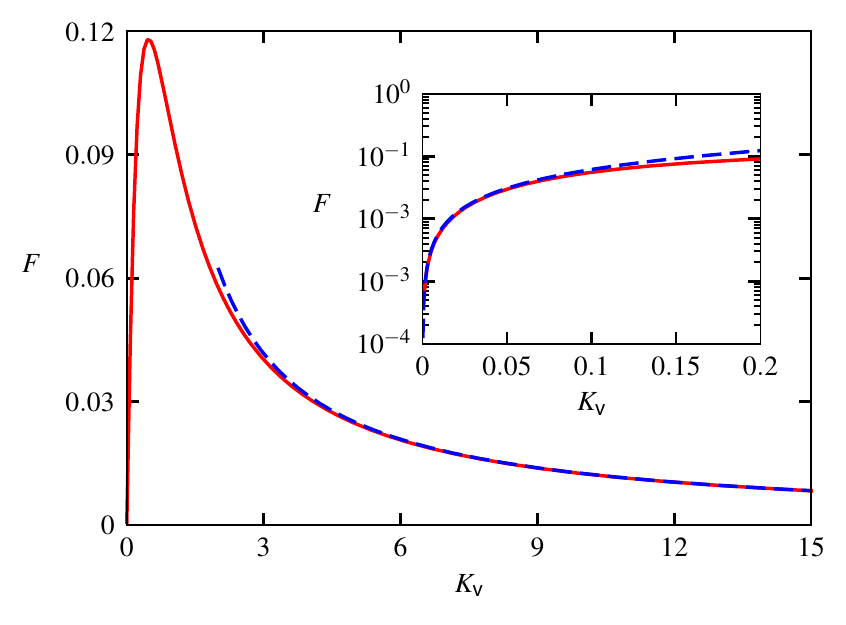}
\caption{\label{fig:vis} Numerical (solid line) and large $K_\text{v}$ asymptotic (dashed line) solution to the propulsive force as a function of $K_\text{v}$ when $b=1$ in viscous fluids. Inset: small $K_\text{v}$ asymptotic results (dashed line) compared with the numerical solution.}
\end{figure}

\section{Propulsive Characteristics}
\label{sec:character_prop}
\subsection{Viscous fluids}
\label{subsec:viscous}
We first discuss the propulsive characteristics of the flexible flapper in a viscous fluid in order to then clearly compare and contrast with propulsion in granular media. In viscous RFT the force density is linearly proportional to the local velocity vector. This linearity permits a simple description of this dynamical system in a viscous fluid, with the (dimensionless) governing equation for $\theta(t)$ given by \cite{peng2016softmatter},
\begin{align}
K_\text{v} \theta + \frac{1}{3}\dot{\theta}+\frac{1}{2}b \cos\theta\cos t=0, \label{eqn:governV}
\end{align}
where $K_\text{v} = \kappa/L^3\xi_\perp\omega$ indicates the ratio of characteristic viscous and elastic forces. One can show that the dimensionless propulsive force (scaled by $L^2\xi_\perp \omega$) is given by
\begin{align}\label{eq:prop_vis}
F = \frac{1}{2}\left< -\dot{\theta} \sin\theta -b\frac{\gamma-1}{\gamma}\cos t\sin 2\theta \right>.
\end{align}
Once the motion is at steady state the term $-\dot{\theta}\sin\theta$ averages to zero as the motion is periodic. We then see that drag anisotropy, $\gamma\ne1$, is required to achieve non-zero propulsion. Furthermore we expect an invariance of the propulsive force under a reversal of the flapping $b\rightarrow -b$ given the symmetry of the steady-state motion of the flapper.

One can easily solve the nonlinear ode for $\theta$ numerically, and integrate the viscous force to obtain the propulsive thrust. As an example, we present in Fig.~\ref{fig:vis} the numerical solution to the propulsive force in a viscous fluid as a function of the spring constant $K_\text{v}$ when the dimensionless actuation amplitude $b=1$. The numerical result (red solid line) matches very well with the large $K_\text{v}$ asymptotic solution (blue dashed line) as given by Eq.~(\ref{eq:largeKvisProp}). We note that for a given spring constant $K_\text{v}$, the propulsive force in general increases as the amplitude $b$ increases, but there is an optimal finite value of the spring constant for a given amplitude which maximizes propulsion. The optimal propulsive force can be readily predicted for small amplitude forcing, $b\ll 1$, to be at $K_\text{v} = 1/3$ to leading order \cite{peng2016softmatter}. To further elucidate the behavior of this torsional spring propeller, we look at the asymptotic regimes where the spring is very soft $K_\text{v} \ll 1$ and when the spring is very stiff $K_\text{v}\gg 1$. 

\begin{figure*}[!ht]
\centering
\includegraphics[width = .9\textwidth]{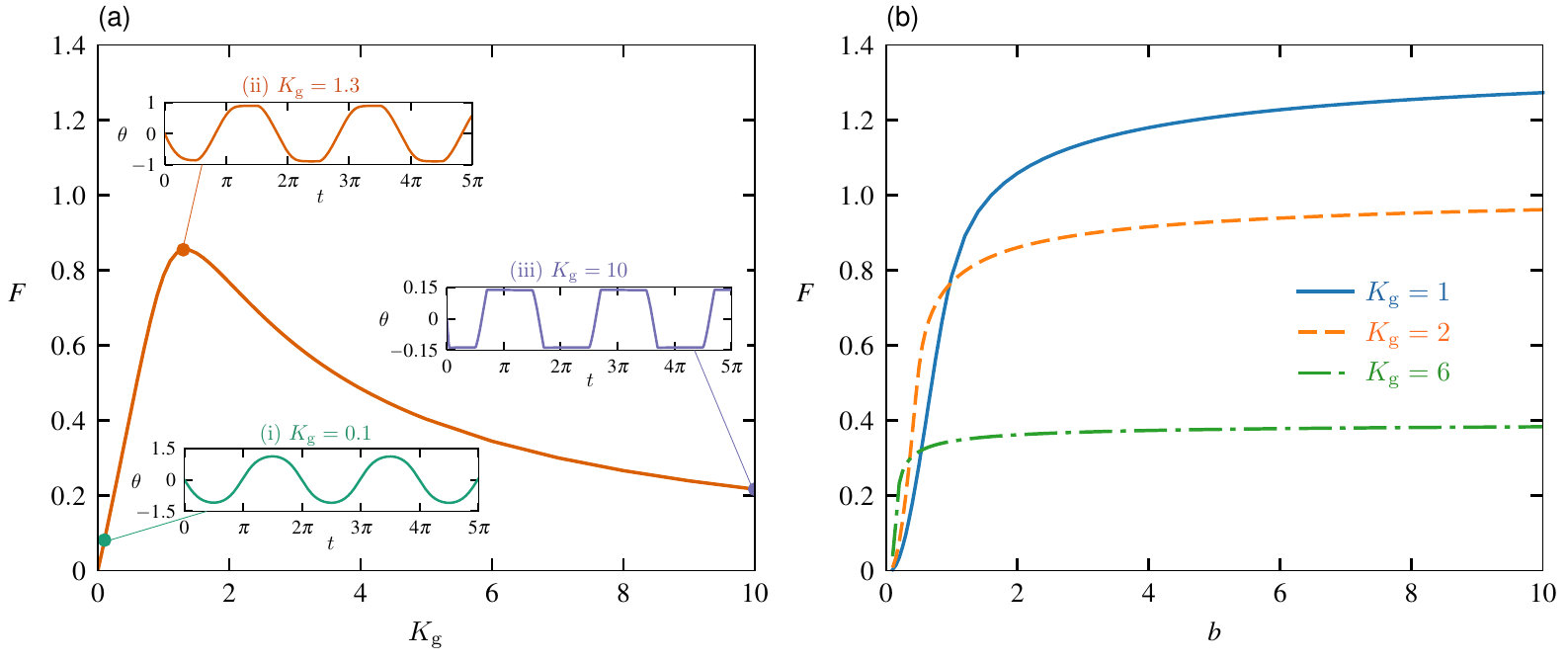}
\caption{\label{fig:prop_finite} (a) Numerical results of the propulsive force in granular media as a function of spring constant $K_\text{g}$ for a dimensionless actuation amplitude of unity ($b=1$); insets: time evolution of the tangent angle $\theta(t)$ for different values of dimensionless spring constants $K_\text{g}$. (b) Propulsive force as a function of the dimensionless actuation amplitude $b$ for different values of $K_\text{g}$. }
\end{figure*}

When the torsional spring is very stiff (the weakly flexible regime) we assume a regular perturbation expansion for the angle $\theta = (1/K_\text{v})\theta_1 + (1/K_\text{v}^2)\theta_2+ \mathcal{O}(1/K_\text{v}^3)$ and find that 
\begin{align}\label{eq:theta_vis}
\theta_1 = -\frac{1}{2}b\cos t,\quad \theta_2 = -\frac{1}{6}b\sin t.
\end{align}
The leading order dimensionless propulsive force is then given by
\begin{align}\label{eq:largeKvisProp}
F = \frac{(\gamma-1)b^2}{4\gamma} \frac{1}{K_\text{v}}+ \mathcal{O}\left(\frac{1}{K_\text{v}^3}\right).
\end{align}
We see that flexibility is necessary for propulsion, as a completely rigid rod, $K_\text{v}\rightarrow \infty$, generates no propulsive force while the propulsive force is inversely proportional to $K_\text{v}$ to leading order. We also confirm that drag anisotropy ($\gamma \neq 1$) is required for propulsion and the leading order term is even in the amplitude as expected.

Conversely, for the case when the torsional spring is instead very weak, $K_\text{v} \ll 1$, we assume that $\theta = \theta_0 + K_\text{v} \theta_1 + \mathcal{O}(K_\text{v}^2)$ and find that
\begin{align}
\theta_0 &= -2 \arctan\left[\tanh \left(\frac{3}{4}b\sin t \right)\right],\\
\theta_1 &= \sech \left( \frac{3}{2} b \sin t \right) \int_0^t h(t^\prime) \text{d} t^\prime,
\end{align}
where the function 
\begin{align}
h(t) = 6\arctan \left[ \tanh \left( \frac{3}{4} b \sin t \right)\right] \cosh\left(\frac{3}{2}b\sin t \right).
\end{align}
The propulsive force can then be written as
\begin{align}
F = -b \frac{\gamma-1}{\gamma}\Bigr< \theta_1\cos t\cos(2\theta_0)\Bigr> K_\text{v} +\mathcal{O}(K_\text{v}^2).
\end{align}
Notice that a completely flexible hinge, $K_\text{v}= 0$, generates no propulsive force and that the propulsive force is even in the amplitude. The agreement between the small $K_v$ asymptotic solution (blue dashed line) and the numerical result is shown in the inset of Fig.~\ref{fig:vis}.

\subsection{Granular media}
In order to obtain a broad understanding of the propulsive characteristics of this elastic propeller in granular media, we first solve the nonlinear governing equation \eqref{eqn:governing} in granular media numerically using a fourth-order Runge-Kutta method coupled with Brent's method for root finding at each time step.

In Fig.~\ref{fig:prop_finite}a, we present the propulsive force as a function of spring constant $K_\text{g}$ for a dimensionless actuation amplitude $b=1$. We observe a non-monotonic variation of the propulsive force as a function of $K_\text{g}$. The maximum propulsive force is achieved when $K_\text{g} \approx 1.3$ with $F \approx 0.85$. 

In the insets of Fig.~\ref{fig:prop_finite}a, we show the time evolution of $\theta$ in granular media for various value of $K_\text{g}$. While the response of $\theta$ to a harmonic actuation in viscous fluids always results in a smoothly varying flapper, regardless of $K_\text{v}$ (see Sec.~\ref{subsec:viscous}), the response of $\theta$ in granular media displays distinct features in different regimes of $K_\text{g}$. When the characteristic granular force dominates the elastic force ($K_\text{g} = 0.1$), the evolution of $\theta$ has a smooth response (Fig.~\ref{fig:prop_finite}a inset i). The evolution of the angle gradually approaches that of a square wave when the spring becomes stiffer (as $K_\text{g}$ increases, Fig.~\ref{fig:prop_finite}a inset iii). The optimal propulsion in granular media is achieved with the right balance of elastic and granular forces ($K_\text{g} = 1.3$, Fig.~\ref{fig:prop_finite}a inset ii).

In Fig.~\ref{fig:prop_finite}b, we show the variation of the propulsive force as a function of the actuation amplitude at various values of $K_\text{g}$. In general, for a given spring constant $K_\text{g}$, the propulsive force increases with the actuation amplitude monotonically and levels off to different values at large amplitudes. 

These numerical simulations reveal a rich class of interesting dynamic responses of a flexible propeller in granular media. To understand the physics and characterize the observed behavior, we perform asymptotic analyses in different regimes in the following section.

\section{Asymptotic analysis}
\label{sec:asymp}

\subsection{Weakly flexible asymptotic solution}
We now consider the case where the torsional spring is very stiff (weakly flexible), \textit{i.e.}, $K_\text{g} \gg 1$.
Assuming a regular series expansion, we write $\theta = (1/K_\text{g})\theta_1+(1/K_\text{g}^2)\theta_2+(1/K_\text{g}^3)\theta_3+\mathcal{O}(1/K_\text{g}^4)$. Substituting the series representation of $\theta$ into the governing equation (Eq.~\ref{eqn:governing}), we obtain
\begin{align}
&\theta_1(t) = -\frac{1}{2}\sgn(\cos t)\left(1+\frac{C_S}{C_F}\cos\gamma_0\right),\nonumber\\
&\theta_3(t) = \frac{1}{4}\sgn(\cos t)\left( 1+ \frac{C_S}{C_F}\sin^2\gamma_0\cos\gamma_0\right) \theta_1^2
\end{align}
while $\theta_2 =0$. Note that the equations for $\theta_1$ and $\theta_3$ are algebraic equations (long time solution), which do not admit initial conditions. The expression for leading order propulsive force, $F = (1/K_\text{g})F_1+\mathcal{O}(1/K_\text{g}^2)$, where
\begin{align}
F_1=\frac{1}{2}\frac{C_S}{C_F}\cos\gamma_0 \left( 1+  \frac{C_S}{C_F}\cos\gamma_0\right). \label{eqn:leading}
\end{align}

\begin{figure}[!tb]
\centering
\includegraphics[width = 0.45\textwidth]{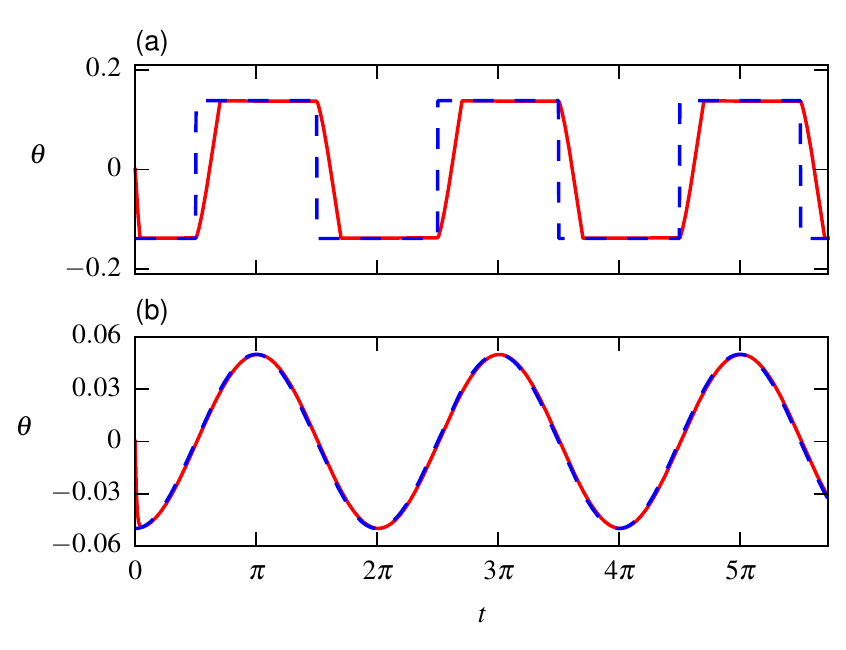}
\caption{\label{fig:theta} Comparison of numerical (solid, red) and leading order asymptotic (dashed, blue) solution to $\theta(t)$ for $K_\text{g} \gg 1$ (or $K_\text{v} \gg 1$ for a viscous fluid) in (a) granular media and (b) viscous fluids for $b=1$. The numerical solutions in the granular and viscous corresponds to $K_\text{g} = 10$ and $K_\text{v} = 10$ respectively. The viscous resistance ratio $\gamma=2$ and the initial condition for numerical simulation is $\theta(t=0)=0$.}
\end{figure}

Unlike in a viscous fluid where the propulsive force for a weakly elastic spring was quadratic in the amplitude of the forcing, in granular media it is independent of the amplitude of actuation to leading order (in $K_\text{g}\gg 1$, Eq.~\ref{eqn:leading}). The evolution of tangent angle from both numerical and asymptotic results in the weakly flexible regime is shown in Fig.~\ref{fig:theta}(a) with $K_\text{g}=10$ and $b=1$. As a comparison, we also present the numerical and asymptotic results for the viscous fluid case in Fig.~\ref{fig:theta}(b). We note that the numerical solution (red solid line) matches well with the asymptotic solution (blue dashed line) in a viscous fluid after a short transience. 

The leading-order large $K_\text{g}$ asymptotic results predict a square-wave response in $\theta(t)$ (blue dashed line, Fig.~\ref{fig:theta}), where the filament maintains a stationary orientation despite varying actuation velocity due to the frictional nature of the granular force model; the torque generated by the velocity-independent granular force balances the elastic force from the torsional spring to give rise to the stationary tangent angle. The jumps in $\theta(t)$ occur at the turning points where the actuation velocity is momentarily zero and the filament changes its direction of motion instantaneously. In the numerical solution, however, the filament takes finite time to adjust its orientation after the turning points and subsequently achieves a stationary configuration again. The discrepancy during these short durations does not affect the propulsive force significantly when $K_\text{g} \gg 1$. We remark that more refined force models beyond the scope of this work are required to resolve the detailed physics in the short durations around the turning points.

\subsection{Steady propulsion and optimal angle}
\label{subsec:steady}

Finally, we calculate in this section an upper limit for the propulsive force that can be generated via this simple torsional spring propeller. Informed by the results in previous sections, far from the turning points on the oscillation path where the actuation velocity changes direction, we expect the filament to reach a stationary configuration with a constant tangent angle.

\begin{figure}[!tb]
\centering
\includegraphics[width = 0.45\textwidth]{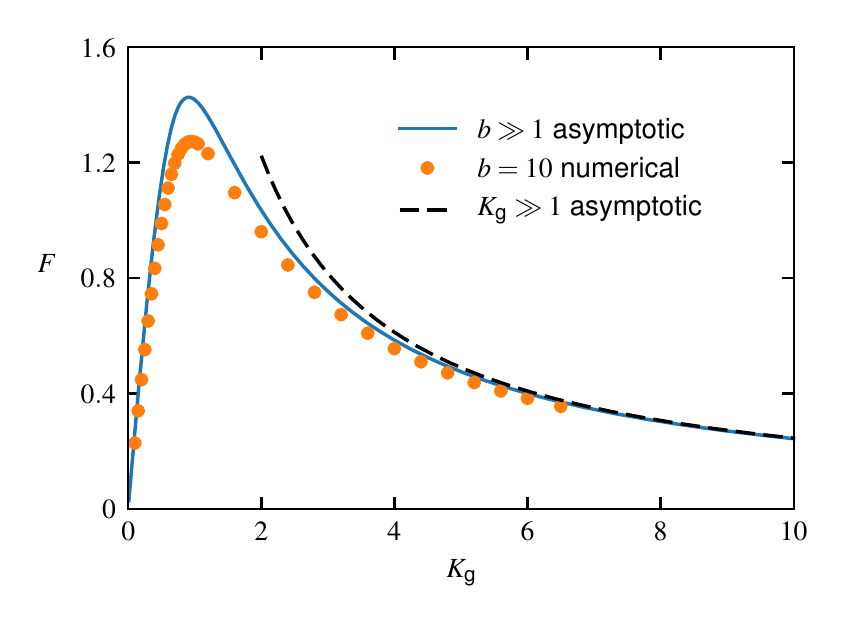}
\caption{\label{fig:largeamp}Comparison of numerical ($b=10$) and large amplitude ($b\gg 1$) asymptotic solution to the propulsive force as a function of spring constant $K_\text{g}$. The black dashed line denotes the large $K_\text{g}$ asymptotic solution.}
\end{figure}

Physically, we can expect an angle which maximizes the propulsive force of this stationary configuration, since no propulsive force is generated when $\theta$ is zero or $\pm\pi/2$. The propulsive force generated with this optimal angle represents an upper limit on the possible propulsive force because any deviation from this optimal angle, albeit inevitable for practical periodic motion, would be suboptimal. Here we calculate this upper limit by enforcing zero angular velocity ($\dot{\theta}=0$), in which case the filament translates in the direction $\hat{\bu} = \sgn(\cos t)\be_y$ while $\hat{\bu}\cdot\bn =\sgn(\cos t) \cos \theta$. In this case \eqref{eqn:governing} simplifies to
\begin{align}
K_\text{g}\theta + \frac{1}{2}\sgn(\cos t)\cos\theta\left(1+\frac{C_S}{C_F\sqrt{\cos^2\theta+\tan^2\gamma_0}}\right)=0.
\end{align}  
Upon assuming the form $\theta = -\Theta ~\sgn(\cos t)$, we obtain
\begin{align}\label{eq:Theta}
K_\text{g}\Theta - \frac{1}{2}\cos\Theta\left(1+\frac{C_S}{C_F\sqrt{\cos^2\Theta+\tan^2\gamma_0}}\right)=0,
\end{align}
while the propulsive force is given by
\begin{align}
F = \sin\Theta(2K_\text{g}\Theta-\cos\Theta).
\end{align}
In this steady limit, we determine the optimal angle as
\begin{align}
\Theta_\text{opt} = \arccos \frac{\sqrt{\sin\gamma_0 }}{\sin\frac{\gamma_0 }{2}+\cos \frac{\gamma_0 }{2}},
\end{align}
which gives rise to a maximum propulsive force of 
\begin{align}
F_\text{max} = \frac{C_S\sqrt{\sin\gamma_0}} { C_F \left( \sin\frac{\gamma_0 }{2}+\cos \frac{\gamma_0 }{2}\right) \sqrt{\sec \gamma_0 \tan\gamma_0(1+\sin\gamma_0)}}
\end{align}
while the corresponding spring constant $K_\text{opt}$ can be obtained using Eq.~(\ref{eq:Theta}). The maximum propulsive thrust obtained in this regime is given by $F_\text{max}\approx 1.427$ with $\Theta_\text{opt} \approx 64^\circ$ and $K_\text{opt}\approx 0.91$ (see blue solid line, Fig.~\ref{fig:largeamp}). We note that the optimal angle can also be recovered from a scaling analysis that maximizes the horizontal propulsive force from a vertical velocity of an element of the filament \cite{peng2016pof}.

As a remark, this steady propulsion limit is also equivalent to the large amplitude asymptotic limit ($b \gg 1$), where the filament translates with the stationary configuration without sampling the turning points. In other words, the leading order solution from a regular perturbation expansion in $1/b$ yields the same results obtained above. In Fig.~\ref{fig:largeamp}, we compare the numerical results for $b=10$ (dots) with the $K_\text{g} \gg 1$ (dashed line) and $b\gg 1$ (solid line) asymptotic results. The large-amplitude asymptotic limit represented by solid blue line should be interpreted as the upper limit of the propulsive force achievable by the simple torsional spring propeller mechanism in granular media. We see that a dimensionless amplitude of $b=10$ already yields close-to-maximum propulsive performance.\\

\section{Conclusion}
\label{sec:conclude}
In this paper, we have presented a combined analytical and numerical investigation to probe the effects of flexibility on propulsion in granular media. While previous works have considered propellers in granular media with prescribed strokes or deformations, the strokes of the flexible propeller considered in this work result from the interaction of elastic and granular forces and we investigated how this can be exploited to maximize propulsive force generation. As a first model, we considered a torsional spring flapper consisting of a rigid filament connected to a torsional spring as a reduced order model that accounts for flexibility. Actuation of the tethered end in a granular medium lead to interesting dynamic features distinct from those in a viscous fluid.

Due to the frictional nature of the granular force model, where the force is independent of velocity, given a harmonic actuation, the filament can develop a non-smooth square-wave-like response in its tangent angle when the elastic forces dominate (large $K_\text{g}$) whereas the response is smoothly varying when the torsional spring is flexible (small $K_\text{g}$). We explained the physics underlying this behavior and derived analytical asymptotic formulae capturing the essential quantitative features of the torsional spring flapper, including the optimal balance of elastic and granular force ($K_\text{opt}$) for maximizing the propulsive force generation. This work serves as a first step to reveal interesting propulsive behaviors that can result from the interplay between body flexibility and its surrounding granular media. The results may be useful for the development of synthetic locomotive systems that exploit flexibility to move through complex terrestrial media.

\bibliography{reference}

\end{document}